\documentclass[journal]{IEEEtran} 
\IEEEoverridecommandlockouts
\usepackage{amsmath,amssymb,amsfonts}
\usepackage{algorithm}
\usepackage{algpseudocode}
\usepackage{graphicx}
\usepackage{textcomp}
\def\BibTeX{{\rm B\kern-.05em{\sc i\kern-.025em b}\kern-.08em
    T\kern-.1667em\lower.7ex\hbox{E}\kern-.125emX}}
\begin{document}

\title{Hybrid Dual Mean-Teacher Network With Double-Uncertainty Guidance for Semi-Supervised Segmentation of MRI Scans}
\author{Jiayi Zhu, Bart Bolsterlee, Brian V. Y. Chow, Yang Song, \IEEEmembership{Member, IEEE}, Erik Meijering, \IEEEmembership{Fellow, IEEE}
\thanks{This work was undertaken with the assistance of resources from the National Computational Infrastructure Australia, a National Collaborative Research Infrastructure Strategy (NCRIS) enabled capability supported by the Australian Government. The authors acknowledge the facilities and scientific and technical assistance of NeuRA Imaging, a node of the National Imaging Facility, an NCRIS capability. This work was supported by an Australian Government Research Training Program Scholarship. The data used in this study were derived from a study funded by the Australian National Health and Medical Research Council (APP1156394). (Corresponding author: Jiayi Zhu.)}
\thanks{Jiayi Zhu, Yang Song and Erik Meijering are with the School of Computer Science and Engineering, University of New South Wales, Sydney, NSW 2052, Australia (e-mails: \{jiayi.zhu3, yang.song1, erik.meijering\}@unsw.edu.au).}
\thanks{Jiayi Zhu, Bart Bolsterlee and Brian V. Y. Chow are with Neuroscience Research Australia (NeuRA), Randwick, NSW 2031, Australia (e-mails: \{b.bolsterlee, b.chow\}@neura.edu.au).}
\thanks{Bart Bolsterlee is also with the Graduate School of Biomedical Engineering, University of New South Wales, Sydney, NSW 2052, Australia}
\thanks{Brian V. Y. Chow is also with the School of Biomedical Sciences, University of New South Wales, Sydney, NSW 2052, Australia}}

\maketitle

\begin{abstract}
Semi-supervised learning has made significant progress in medical image segmentation. However, existing methods primarily utilize information acquired from a single dimensionality (2D/3D), resulting in sub-optimal performance on challenging data, such as magnetic resonance imaging (MRI) scans with multiple objects and highly anisotropic resolution. To address this issue, we present a Hybrid Dual Mean-Teacher (HD-Teacher) model with hybrid, semi-supervised, and multi-task learning to achieve highly effective semi-supervised segmentation. HD-Teacher employs a 2D and a 3D mean-teacher network to produce segmentation labels and signed distance fields from the hybrid information captured in both dimensionalities. This hybrid learning mechanism allows HD-Teacher to combine the `best of both worlds', utilizing features extracted from either 2D, 3D, or both dimensions to produce outputs as it sees fit. Outputs from 2D and 3D teacher models are also dynamically combined, based on their individual uncertainty scores, into a single hybrid prediction, where the hybrid uncertainty is estimated. We then propose a hybrid regularization module to encourage both student models to produce results close to the uncertainty-weighted hybrid prediction. The hybrid uncertainty suppresses unreliable knowledge in the hybrid prediction, leaving only useful information to improve network performance further. Extensive experiments of binary and multi-class segmentation conducted on three MRI datasets demonstrate the effectiveness of the proposed framework. Code is available at https://github.com/ThisGame42/Hybrid-Teacher.
\end{abstract}


\section{Introduction}
\IEEEPARstart{S}{emantic} segmentation of medical imaging data, such as magnetic resonance imaging (MRI) scans, have been an integral part of many clinical applications \cite{flair_ref, segmentation_ref_1}. Segmentation is typically performed manually by human experts on a slice-by-slice basis, which is time-consuming. The recent rise of deep neural networks has enabled automatic segmentation in many tasks \cite{nnUNet, UNet, VNet}. However, the need for large amounts of labeled data poses an obstacle to harness the full power of deep segmentation networks. Numerous attempts have been made to alleviate this problem, such as: (1) interactive learning \cite{interaction_1, interaction_2, interaction_3} where human users assess the intermediate outputs from the networks to improve the performance; (2) weakly-supervised learning \cite{weak_supervision_1, weak_supervision_2} where the networks learn from easy-to-acquire coarse annotations; and (3) semi-supervised learning \cite{u1, u2} where the networks learn from a small set of labeled data and a larger set of unlabeled data. We focus on semi-supervised networks in this work because they perform well while requiring minimum human annotations and interventions, making them suitable for integration into clinical and research settings \cite{flair_ref_2}.

Recently proposed semi-supervised methods often aim to enforce consistency between the network outputs produced on labeled and unlabeled data to regularize the network training. Examples include the SASSNet \cite{SASSNet}, where a multi-task segmentation network producing segmentation labels and signed distance fields (SDF) \cite{SDF, SDF_2} is regularized by a discriminator to produce consistent SDF on labeled and unlabeled data. The mean-teacher network \cite{mean_teacher} imposes data-level and model-level consistency by training a student model, using its weights to update an ensemble model, and enforcing mutually-agreed outputs on data under different perturbations. Consistency regularization is further explored in \cite{MTL1, MTL2}, which combine multi-task learning with the mean-teacher network and propose a triple-uncertainty-guided mean-teacher approach imposing uncertainty-guided data-level, model-level, and task-level consistency.

Existing semi-supervised networks typically operate either on 2D or on 3D images, not both. However, this single-mode learning approach is limited. As pointed out in \cite{h_denseunet, hybridnet1, hybridnet2}, 2D networks do not probe along the third dimension of volumetric data, whereas the high computational cost and memory constraints limit the depth of 3D networks. Furthermore, 3D networks struggle to extract features efficiently from data with high spatial anisotropy (e.g., volumetric data with a voxel size of $0.4 \times 0.4 \times 5$ mm) due to the large spacing between the slices \cite{hybridnet1}. 2.5D segmentation networks such as \cite{2.5D} have been proposed to alleviate this issue. They stack adjacent slices from different orientations together and use 2D operations such as convolution to process them. Likewise, hybrid networks augment 3D features with their 2D counterparts to enhance the network's feature learning capability and have demonstrated promising performance if trained on sufficient data \cite{h_denseunet, hybridnet1}.

Inspired by recent progress in hybrid and multi-task learning, we propose a hybrid mean-teacher network which we name Hybrid Dual Mean-Teacher (HD-Teacher), to achieve efficient feature learning with minimal data annotation. Different from existing mean-teacher networks such as \cite{MTL1, MTL2}, the proposed HD-Teacher simultaneously utilizes \textit{two} sets of mean-teacher networks, one 2D and one 3D, to extract hybrid features and produce segmentation labels, SDF predictions and their uncertainties. SDF measures the distance between a given pixel and the nearest boundary of foreground objects and is predicted as a secondary task to model the shape information into the HD-Teacher, allowing for more accurate segmentation results with refined surfaces. To address the issue that a 2D/3D mean-teacher network may struggle to extract adequate information from a single dimensionality, we first use segmentation probability maps from the 2D mean-teacher network to distill contextual information to be used by the 3D mean-teacher network. Then, features produced by the 2D mean-teacher network and its 3D counterpart are fused to construct hybrid features, which train both mean-teacher networks with information extracted from 2D and 3D spaces. Next, HD-Teacher dynamically fuses outputs from 2D and 3D teacher models into a hybrid prediction using individually estimated uncertainty and estimates the overall hybrid uncertainty from the result. The hybrid uncertainty suppresses unreliable parts in the hybrid prediction and ensures that both students are jointly regularized by trustworthy hybrid information from both dimensions. As such, HD-Teacher effectively addresses the weakness of existing mean-teacher networks and, as a result, demonstrates significant performance improvement.

Our contributions can be summarized as follows:
\begin{itemize}
    \item We propose to integrate semi-supervised, multi-task and hybrid feature learning, and develop a novel hybrid dual mean-teacher network for semi-supervised segmentation of medical images. The proposed framework utilizes a 2D and a 3D component to learn hybrid features that are rich in semantic and geometric information.
    \item We propose to dynamically merge outputs from 2D and 3D teacher models into a single hybrid prediction and estimate the overall hybrid uncertainty of the entire framework from the result. We then use the hybrid uncertainty to re-weigh the hybrid prediction for regularization, enabling both student models to learn from reliable hybrid knowledge by encouraging them to produce results close to the calibrated hybrid prediction. 
    \item We demonstrate that the proposed framework offers a significant performance improvement over state-of-the-art semi-supervised methods through extensive experiments involving binary and multi-class segmentation of three MRI datasets, including both anisotropic and isotropic data.
\end{itemize}

\section{Related Work}

\subsection{Semi-Supervised Learning for Segmentation}
Semi-supervised segmentation networks aim to leverage labeled and unlabeled data to improve segmentation accuracy. One popular category of semi-supervised segmentation is methods producing pseudo labels. Examples include \cite{p_label}, where a neural network produces pseudo labels and updates its parameters in an alternating manner. Adversarial learning is another popular technique to mine useful information from unlabeled data. For instance, a discriminator network can encourage similar SDF predictions to be produced on labeled and unlabeled data \cite{SASSNet}.

Lately, methods enforcing consistency have been gaining popularity due to their performance. Typically, consistency is enforced at data, model, or task level. Data-level consistency models, such as the $\Pi$-model proposed in \cite{data_con_1}, aim to produce consistent predictions on data with variable noise levels using temporal ensembles of outputs from previous epochs for regularization. Networks with task-level consistency such as \cite{DTC} encourage consistent predictions, regardless of the task, to be made on the same input. The mean-teacher network \cite{mean_teacher} improves the $\Pi$-model by enforcing consistent predictions between the student model and its ensemble, an exponential moving average (EMA) of the student model, thereby achieving data and model-level consistency. One issue with the mean-teacher is that the ensemble model may produce unreliable results. Thus, uncertainty estimation \cite{u1, u2, u3} is integrated with the mean-teacher network to facilitate more reliable knowledge transfer from the ensemble, resulting in more accurate segmentation labels than competing methods. Uncertainty-guided mean-teacher networks have been further improved by using triple-uncertainty in \cite{MTL1, MTL2}, where data, model and task-level consistencies are combined via multi-task learning to exploit unlabeled data with higher efficiency.

Unlike existing mean-teacher networks operating on images of only a single dimensionality, we propose a hybrid setup wherein a set of 2D and 3D mean-teacher networks extract, exchange, and merge 2D and 3D information toward more accurate segmentation labels and uncertainty estimation and enable cross-dimension consistency regularization.

\begin{figure*}
\centerline{\includegraphics[width=\textwidth]{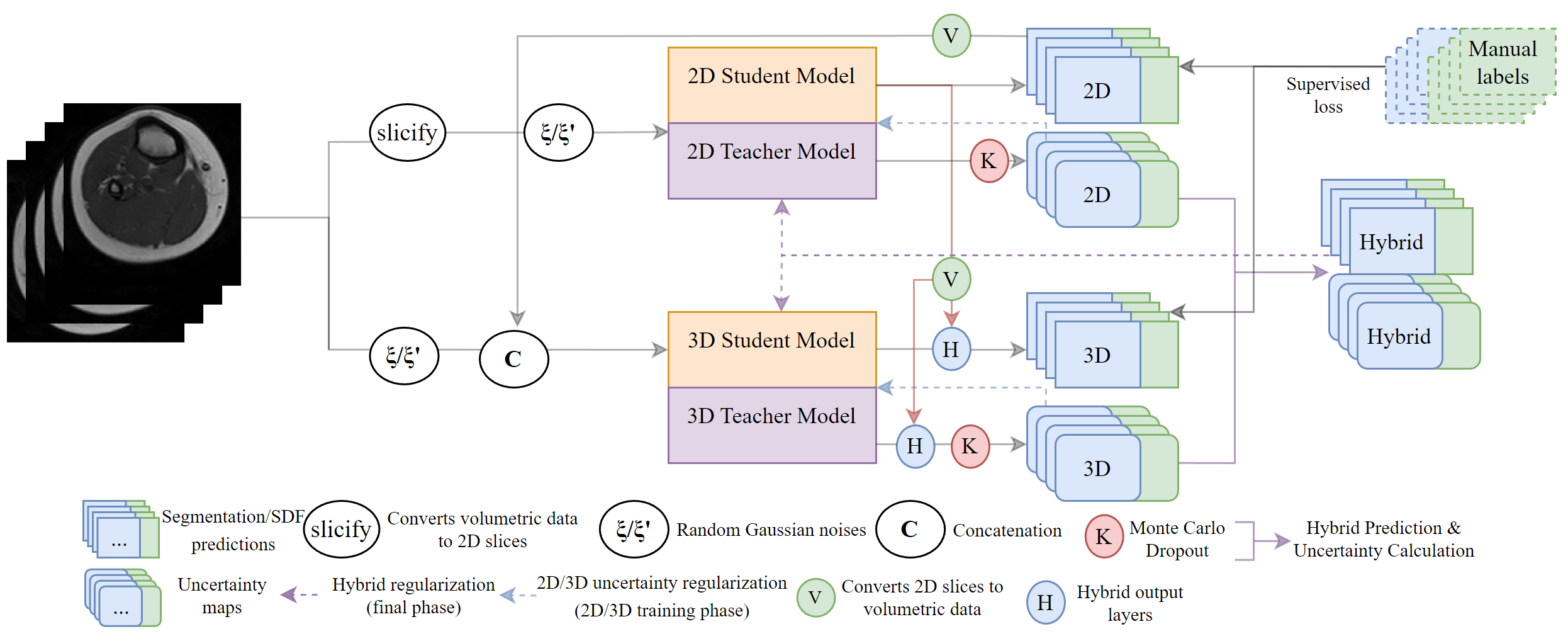}}
\caption{Schematic view of the proposed HD-Teacher, where a 2D and a 3D uncertainty-guided multi-task mean-teacher network work in tandem to produce segmentation and SDF predictions using hybrid features. Each mean-teacher network has a student model, trained using stochastic gradient descent, and a teacher model, which is an ensemble of the student model and provides uncertainty-weighted regularization. The 2D student model is first trained and regularized by predictions and uncertainties from the 2D teacher model. Then, its 3D counterpart is augmented by 2D features and segmentation maps while being trained and regularized by outputs and uncertainties from the 3D teacher model. Finally, both student models are jointly fine-tuned with hybrid prediction and uncertainty calculated from a weighted combination of outputs from 2D and 3D teacher models. The complete training procedure is summarized in Algorithm \ref{algorithm1}.}
\label{fig1}
\end{figure*}

\subsection{Feature Learning for Segmentation}
Numerous attempts have been made to include inter-slice information and spatial contexts in 2D neural networks. A notable approach is 2.5D networks \cite{2.5D, 2.5D_1}, which refer to frameworks using 2D convolutional neural networks (CNNs) with adjacent slices, potentially from different planes and orientations, as input. 2.5D networks can also be implemented by combining 2D and 3D operations in a single model where higher layers in the encoder and decoder employ 2D operations, and lower layers use 3D ones to approximate a balanced receptive field along each axis \cite{2.5D_2}. 2.5D networks, regardless of their implementations, outperform traditional 2D/3D networks, demonstrating the importance of learning features from multiple sources and orientations.

To further improve the feature learning capability of neural networks, hybrid feature learning merges detailed intra-slice/2D features with 3D contexts captured by inter-slice/3D features to improve model performance. In addition, hybrid learning enables deep and efficient 3D networks \cite{h_denseunet}, and extracts features more effectively from spatially anisotropic data \cite{hybridnet2}. We further investigated hybrid learning in \cite{hybridnet1} with spatial and channel-wise attention \cite{scse} to boost model performance. Likewise, a hybrid 2D/3D CNN architecture is proposed in \cite{hybridnet3} to fuse 2D and 3D features in the downsampling phase for segmenting chronic stroke lesions. 2D and 3D features from 2D and 3D CNNs are merged in \cite{hybridnet4, hybridnet_5} to segment the lung lobe and tumors, respectively.

The proposed HD-Teacher improves upon existing hybrid networks in (1) practicality by reducing the requirement of data annotation via semi-supervised training, (2) generalizability by employing uncertainty-guided regularization, and (3) performance through multi-task learning to refine shapes of segmented objects.

\subsection{Multi-Task Learning for Segmentation}
Multi-task learning is a technique wherein multiple relevant tasks are learned in tandem, and shared knowledge is utilized to improve generalization. In the case of neural networks, knowledge of different tasks is typically propagated via soft or hard parameter sharing. In soft parameter sharing, different models are trained for different tasks, and regularization is used to minimize the difference between their model weights \cite{soft_1, soft_2}. In hard parameter sharing, several output branches are integrated into a single model, producing results for different tasks using shared features from lower layers \cite{hard_1}.

Multi-task learning has enabled significant performance improvement in the task of medical image segmentation. In \cite{multi_learning_1}, a single CNN takes different imaging modalities as input and learns common features for the segmentation task. Segmentation and classification of tumors from breast ultrasound images are jointly performed in \cite{multi_learning_2} using feature maps from a shared encoder. Joint glioma segmentation and isocitrate dehydrogenase are integrated into a single framework in \cite{multi_learning_3} using correlated features from a CNN-Transformer encoder. A shape-refined segmentation framework with atlas propagation is proposed in \cite{multi_learning_4} to learn segmentation and landmark location simultaneously. Shape information of foreground objects is utilized in \cite{multi_learning_5, multi_learning_6} to enhance the details around the edges of the predicted segmentation labels.

Motivated by these existing works, we employ the hard parameter-sharing strategy by utilizing a shared encoder for segmentation and SDF tasks, ensuring that learned features contain semantic and geometric shape information while keeping HD-Teacher compact.

\section{Methods}

The mean-teacher network refers to a framework where a student model is trained, and the EMA of its weights is used to update an ensemble model called the teacher model. Our framework integrates hybrid learning into the mean-teacher architecture, utilizing one 2D and one 3D mean-teacher network to extract hybrid features from both dimensionalities for the segmentation and SDF prediction tasks (Fig. \ref{fig1}). This hybrid setup overcomes issues associated with single-dimension learning by (1) appending contextual information to the 3D input volume and (2) fusing 2D and 3D features weighted by their uncertainties,  allowing for more efficient and more accurate predictions (described in upcoming sections). It trains in stages: the 2D and 3D mean-teacher networks are trained separately before they are jointly fine-tuned (Algorithm \ref{algorithm1}). The training dataset is divided into the labeled set $D^\text{l} =\{X^{i}, Y^{i}, Z^{i}\}_{i=1}^{N}$ and unlabeled set $D^\text{u} = {\{X^{i}\}}_{i=N+1}^{N+M}$, where $N$ and $M$ denote the number of labeled and unlabeled samples, and $X$ and $Y$ represent the input image and the corresponding reference labels, respectively. $Z$ denotes the ground-truth SDF which is used for multi-task learning and is calculated using the SDF function proposed in \cite{SDF}. We set $M \ge 4N$ across all our experiments.

\subsection{2D Mean-Teacher Network} \label{mean_teacher_network_explained}
The student model in the 2D mean-teacher network follows the design of the popular U-Net \cite{UNet}, except for (1) the initial number of feature maps, which is set to 32 to match the dimension of the 3D mean-teacher network for later feature fusion and (2) it has one encoder, which is shared between its two decoders: one for segmentation and the other for predicting SDF maps. This multi-task learning paradigm enforces the encoder to extract semantic features with rich geometric shape information, allowing both decoders to succeed in their tasks. It also implicitly establishes cross-task guidance, enabling each decoder to improve its output with additional semantic/shape information.

Given a batch of 2D input images, $X^\text{2D} \in \mathbb{R}^{b \times c \times w \times h}$, where $b$, $c$, $w$ and $h$ denote batch size, number of channels, and image width and height, respectively, the 2D student model produces segmentation labels, SDF predictions, and feature maps learned by the last block of both decoders:
\begin{equation} \label{eq1}
\hat{Y}^\text{s2d}, \hat{Z}^\text{s2d}, F^\text{s2d}_\text{seg}, F^\text{s2d}_\text{sdf} = f^\text{s2d}(X^\text{2D}; \theta^\text{s2d}, \xi)
\end{equation}
where $\hat{Y}^\text{s2d}$ and $\hat{Z}^\text{s2d}$ denote the predicted segmentation probability and SDF maps activated by the softmax and tanh function, respectively. $F^\text{s2d}_\text{seg}$ and $F^\text{s2d}_\text{sdf}$ represent feature maps learned by the last block of segmentation and SDF decoders. These feature maps are later fused with their 3D counterparts to generate the final segmentation and SDF predictions (Section \ref{3D_nets}). $f^\text{s2d}$ indicates both decoders in the student model and $\theta^\text{s2d}$ their parameters. $\xi$ is random Gaussian noise added to $X^\text{2D}$ to enhance the robustness of the networks. We omit the shared encoder and its parameters for simplicity.

The teacher model is identical to the student model except for an additional dropout layer \cite{dropout} inserted after its encoder's last layer. Therefore, the teacher model has identical outputs and utilizes them to regularize the student model. To ensure a reliable regularization process,  the teacher model measures the uncertainty from its predictions to filter out unreliable ones. We follow \cite{u1, u2} and estimate the uncertainties in the teacher's outputs using Monte Carlo Dropout \cite{mc_dropout}. More specifically, in each training step, the teacher model performs $K$ forward passes, resulting in $K$ intermediate segmentation probability maps $\{\hat{Y}^\text{t2d}_{j}\}_{j=1}^{K}$ and SDF predictions $\{\hat{Z}^\text{t2d}_{j}\}_{j=1}^{K}$ being produced under random dropouts and noise perturbations (i.e., random Gaussian noise added to the input image). The uncertainty of each intermediate segmentation probability map $U_{j}^\text{seg2d}$ is then estimated as the entropy:
\begin{equation} \label{eq2}
    U_{j}^\text{seg2d} = -\sum_{c \in C} \hat{Y}_{j,c}^\text{t2d}\text{log}_{C}\hat{Y}_{j,c}^\text{t2d}
\end{equation}
where the log function is based on the number of classes $C$ for the segmentation task. The entropy $U_{j}^\text{seg2d} \in [0, 1]$ indicates areas with high uncertainty (large values), allowing the teacher model’s confidence in each intermediate segmentation probability map to be estimated as $\{1- U_{j}^\text{seg2d}\}_{j=1}^{K}$.

We then follow \cite{MTL1, MTL2} to normalize all confidence maps to $[0, 1]$ by stacking them and applying the softmax function, i.e., $\{W\}_{j=1}^{K} = \text{softmax}(\{1 - U_{j}^\text{seg2d}\}_{j=1}^{K})$. $W_{j}$, the $j$th channel of the result, indicates the trustworthiness of the corresponding intermediate segmentation probability map. The final probability map is calculated as a weighted sum of intermediate probability maps and their trustworthiness to assign more weights to more trustworthy intermediate results:
\begin{equation} \label{eq3}
    \hat{Y}^\text{t2d} = \sum_{j=1}^{K}W_{j} \odot \hat{Y}_{j}^\text{t2d}
\end{equation}
where $\odot$ denotes the element-wise multiplication operation. The overall segmentation uncertainty of the teacher model is estimated again as the entropy, i.e., $U^\text{seg2d} = -\sum_{n\in N} \hat{Y}^\text{t2d}_{n}\text{log}_{N}\hat{Y}^\text{t2d}_{n}$. $U^\text{seg2d}$ is later used to suppress unreliable parts in $\hat{Y}^\text{t2d}$ to establish a more effective knowledge transfer to the student model (Section \ref{loss_fn}).

The uncertainty for the SDF prediction task cannot be estimated using the entropy method since SDF prediction is implemented as a regression task with real-valued outputs rather than probabilities. Thus, as per \cite{mc_dropout}, we first obtain the final SDF prediction by taking the average of $K$ intermediate SDF predictions as $\hat{Z}^\text{t2d} = \frac{1}{K}\sum_{j=1}^{K}\hat{Z}^\text{t2d}_{j}$ and estimate the uncertainty as the variance, i.e., $U^\text{sdf2d} = \frac{1}{K}\sum_{j=1}^{K}(\hat{Z}_{j}^\text{t2d} - \hat{Z}^\text{t2d})^{2}$ to regularize the student model with reliable information.

At the end of the 2D training phase, we freeze the weights of both 2D models and proceed to train the 3D mean-teacher network.

\subsection{3D Mean-Teacher Network}\label{3D_nets}
The 3D mean-teacher network is architecturally identical to its 2D counterpart except all layers operate in the 3D space, allowing it to also benefit from the multi-task learning setup. However, 3D models have been shown to perform poorly on anisotropic data due to their inability to extract semantic-rich information from adjacent slices with large gaps in between \cite{h_denseunet, hybridnet1, hybridnet2}. As such, we propose to augment the 3D mean-teacher network with features and segmentation outputs from its 2D equivalent to enhance its learning capability.

Specifically, we first define a function $C$ to convert a batch of 3D volumes $X^\text{3D} \in \mathbb{R}^{b\times c \times d \times w \times h}$ to 2D images $\tilde{X}^\text{2D} \in \mathbb{R}^{bd \times c \times w \times h}$ by stacking the batch dimension $b$ and the depth dimension $d$ together. The inverse transformation of $C$ is denoted as $C^{-1}$. Then, in each training step, we first acquire the 2D outputs and feature maps from the trained 2D student model by replacing $X^\text{2D}$ in Eq. \eqref{eq1} with $\tilde{X}^\text{2D}$. The obtained 2D segmentation outputs are then converted to volumetric shapes using $C^{-1}$ and concatenated with the original volumes $X^\text{3D}$ to form $\tilde{X}^\text{3D}$, which is fed to 3D models to distill contextual information and provide additional guidance. Next, the 3D student model processes $\tilde{X}^\text{3D}$ with its encoder and produces segmentation feature maps $F^\text{s3d}_\text{seg}$ and SDF feature maps $F^\text{s3d}_\text{sdf}$ from its decoders. These feature maps are then fused with their 2D counterparts (in volumetric shapes via $C^{-1}$) using an element-wise summation operation to construct hybrid features. Lastly, the final predictions are generated from hybrid features using $1 \times 1 \times 1$ convolution operations. We add random Gaussian noise to $\tilde{X}^\text{3D}$ and $\tilde{X}^\text{2D}$ to further enhance the robustness of HD-Teacher.


The forward pass of the 3D teacher model is identical except for the addition of the Monte Carlo Dropout, which performs $K$ forward passes to estimate the 3D segmentation uncertainty $U^\text{seg3d}$ and SDF uncertainty $U^\text{sdf3d}$ identically to its 2D counterpart (see Eq. \eqref{eq2} and onward). The segmentation and SDF outputs and their corresponding feature maps of the 3D teacher model are denoted as $\{\hat{Y}^\text{t3d}_{j}\}_{j=1}^{K}$, $\{\hat{Z}^\text{t3d}_{j}\}_{j=1}^{K}$, $F^\text{s3d}_\text{seg}$, and $F^\text{s3d}_\text{sdf}$, respectively.

\subsection{Hybrid Prediction and Uncertainty Estimation}
After training the 3D mean-teacher network with a fixed 2D student model, we `unfix' its weights and jointly fine-tune both 2D and 3D mean-teacher networks to improve performance with a hybrid co-training paradigm. As part of the process, we propose to dynamically merge outputs from 2D and 3D teacher models into a single hybrid prediction based on their individual uncertainty scores and estimate the hybrid uncertainty, which assesses the confidence level of the entire framework, from the result. The estimated hybrid uncertainty enables (1) re-weighing of the hybrid prediction to suppress questionable information from the perspective of both dimensionalities and (2) cross-dimensionality hybrid consistency to be imposed between the calibrated hybrid prediction and both student models for further performance improvement.

Given 2D and 3D intermediate segmentations and SDF predictions from their respective teacher models $\{\hat{Y}^\text{t2d}_{j}\}_{j=1}^{K}$, $\{\hat{Z}^\text{t2d}_{j}\}_{j=1}^{K}$, $\{\hat{Y}^\text{t3d}_{j}\}_{j=1}^{K}$, and $\{\hat{Z}^\text{t3d}_{j}\}_{j=1}^{K}$, we first generate the final hybrid segmentation prediction $\hat{Y}^\text{h}$ as an uncertainty-weighted combination of 2D and 3D intermediate segmentation outputs by:
\begin{equation} \label{eq5}
    \begin{gathered}
        \{\hat{Y}_{j}^\text{h}\}_{j=1}^{2K} = \text{Con}\left(C^{-1}\left(\{\hat{Y}_{j}^\text{t2d}\}_{j=1}^{K}\right), \{\hat{Y}_{j}^\text{t3d}\}_{j=1}^{K}\right) \\
        \{U_{j}^\text{h}\}_{j=1}^{2K} = \left\{-\sum_{n \in N} \hat{Y}^\text{h}_{j,n}\text{log}_{N}\hat{Y}_{j,n}^\text{h}\right\}_{j=1}^{2K}\\
        \{W\}_{j=1}^{2K} = \text{softmax}\left(\{1 - U_{j}^\text{h}\}_{j=1}^{2K}\right)\\
        \hat{Y}^\text{h} = \sum_{j=1}^{2K} W_{j} \odot \hat{Y}_{j}^\text{h}
    \end{gathered}
\end{equation}
where Con denotes the concatenation operation and $C^{-1}$ converts 2D tensors to volumetric shapes. $\{W\}_{j=1}^{2K}$ is an entropy-based weight map that reflects the trustworthiness of all (2D and 3D) intermediate segmentation maps at each pixel location. The hybrid segmentation uncertainty is then estimated as the entropy of $\hat{Y}^\text{h}$, i.e., $U^\text{seg}_\text{h} = -\sum_{n \in N} \hat{Y}^\text{h}_{n}\text{log}_{N}\hat{Y}^\text{h}_{n}$. Similarly, the final hybrid SDF prediction is obtained by concatenating all 2D and 3D intermediate SDF predictions together and averaging the result, which is defined as:
\begin{equation} \label{eq6}
    \begin{gathered}
        \{\hat{Z}_{j}^\text{h}\}_{j=1}^{2K} = \text{Con}\left(C^{-1}\left(\{\hat{Z}_{j}^\text{t2d}\}_{j=1}^{K}\right), \{\hat{Z}_{j}^\text{t3d}\}_{j=1}^{K}\right) \\
        \hat{Z^\text{h}} = \frac{1}{2K}\sum_{j=1}^{2K}\hat{Z}_{j}^\text{h}\\
    \end{gathered}
\end{equation}
The hybrid SDF uncertainty is then measured as the variance of $\hat{Z^\text{h}}$, i.e., $U^\text{sdf}_\text{h} = \frac{1}{2K} \sum_{j=1}^{2K}(\hat{Z}^\text{h}_{j} - \hat{Z}^\text{h})^2$.

\subsection{Training Procedures: 2D, 3D, Hybrid} \label{loss_fn}
We optimize both student models with two loss terms: supervised loss $L_\text{s}$ and consistency loss $L_\text{c}$. The supervised loss $L_\text{s}$ is defined as:
\begin{equation} \label{loss_s}
    L_\text{s}(\hat{Y}^\text{s}, Y, \hat{Z}^\text{s}, Z) = \frac{1}{N}\sum_{i=1}^{N}\left(L_\text{seg}(\hat{Y}^\text{s}_{i}, Y_{i}) + L_\text{sdf}(\hat{Z}^\text{s}_{i}, Z_{i})\right)
\end{equation}
where $\hat{Y}^\text{s}$ and $Y$ represent the segmentation probability map produced by a student model and the corresponding reference labels. Likewise, $\hat{Z}^\text{s}$ and $Z$ are the SDF predictions made by a student model and the corresponding ground-truth SDF, respectively. $L_\text{seg}$ is implemented with the Dice loss function \cite{VNet} and $L_\text{sdf}$ uses the mean squared error (MSE). The EMA of the student model's weights is used to update the teacher model at each training iteration with an EMA decay coefficient $\tau$ to control the update rate. Teacher models are expected to produce consistent results as student models on the same input data due to their ensemble nature. We enforce such consistency between student and teacher models with $L_\text{c}$:
\begin{equation} \label{loss_c}
    \begin{gathered}
        L_\text{c}(\hat{Y}^\text{s}, \hat{Y}^\text{t}, \hat{Z}^\text{s}, \hat{Z}^\text{t}, U^\text{seg}, U^\text{sdf}) = \\ L_\text{c}^\text{seg}(\hat{Y}^\text{s}, \hat{Y}^\text{t}, U^\text{seg}) + L_\text{c}^\text{sdf}(\hat{Z}^\text{s}, \hat{Z}^\text{t}, U^\text{sdf})\\
        L_\text{c}^\text{seg} = \frac{1}{N+M} \sum_{i=1}^{N+M}\text{exp}(-U^\text{seg}_{i}) \odot (\hat{Y}_{i}^\text{s} - \hat{Y}_{i}^\text{t})^2\\
        L_\text{c}^\text{sdf} = \frac{1}{N+M} \sum_{i=1}^{N+M}\text{exp}(-U^\text{sdf}_{i}) \odot (\hat{Z}_{i}^\text{s} - \hat{Z}_{i}^\text{t})^2
    \end{gathered}
\end{equation}
where $U^\text{seg}$ and $U^\text{sdf}$ are segmentation and SDF uncertainties estimated by a teacher model, and $\hat{Y}^\text{s}, \hat{Z}^\text{s}, \hat{Y}^\text{t}, \hat{Z}^\text{t}$ are outputs from a student and a teacher model, respectively.

As mentioned, the 2D and 3D mean-teacher networks are trained separately before they are jointly fine-tuned for optimal performance (Algorithm \ref{algorithm1}). The 2D student model is first trained using Eqs. (\ref{loss_s}) and (\ref{loss_c}) as:
\begin{equation} \label{eq2d}
    \begin{gathered}
    L_\text{2d} = L_\text{s}(\hat{Y}^\text{s2d}, Y, \hat{Z}^\text{s2d}, Z) + \\
    \lambda^\text{2d}L_\text{c}(\hat{Y}^\text{s2d}, \hat{Y}^\text{t2d}, \hat{Z}^\text{s2d}, \hat{Z}^\text{t2d}, U^\text{seg2d}, U^\text{sdf2d})
    \end{gathered}
\end{equation}
where $\lambda^\text{2d}$ controls the strength of the consistency loss. We then fix the weights of the 2D student model and use it to help train the 3D student model using Eqs. (\ref{loss_s}) and (\ref{loss_c}) as:
\begin{equation} \label{eq3d}
    \begin{gathered}
        L_\text{3d} = L_\text{s}(\hat{Y}^\text{s3d}, Y, \hat{Z}^\text{s3d}, Z) + \\
    \lambda^\text{3d}L_\text{c}(\hat{Y}^\text{s3d}, \hat{Y}^\text{t3d}, \hat{Z}^\text{s3d}, \hat{Z}^\text{t3d}, U^\text{seg3d}, U^\text{sdf3d})
    \end{gathered}
\end{equation}
where $\lambda^\text{3d}$ again balances the contributions of the consistency loss. Finally, the 2D and 3D student models are jointly trained using reference labels and regularized by the hybrid prediction and uncertainty to achieve cross-dimensionality consistency:
\begin{equation} \label{eq10}
    \begin{gathered}
        L_\text{h} = L_\text{h}^\text{3d} + \alpha L_\text{h}^\text{2d}\\
        L_\text{h}^\text{2d} = L_\text{s}(\hat{Y}^\text{s2d}, Y, \hat{Z}^\text{s2d}, Z)  + \lambda^\text{2d}L_\text{c}(\hat{Y}^\text{s2d}, \hat{Y}^\text{h}, \\
        \hat{Z}^\text{s2d}, \hat{Z}^\text{h}, U^\text{seg}_\text{h}, U^\text{sdf}_\text{h})\\
        L_\text{h}^\text{3d} = L_\text{s}(\hat{Y}^\text{s3d}, Y, \hat{Z}^\text{s3d}, Z)  + \lambda^\text{3d}L_\text{c}(\hat{Y}^\text{s3d}, \hat{Y}^\text{h}, \\
        \hat{Z}^\text{s3d}, \hat{Z}^\text{h}, U^\text{seg}_\text{h}, U^\text{sdf}_\text{h})\\
    \end{gathered}
\end{equation}
where $\alpha$ is a hyperparameter set to balance the contributions of the 2D student model.

\begin{algorithm}
\caption{Training procedure of HD-Teacher.}
\begin{algorithmic}
\State $\textbf{Require: }\text{2D and 3D mean-teacher models.}$
\State $\textbf{2D stage:}$
\State $\text{1. Train the 2D student model with the combined 2D loss}$
\State $\text{$L_\text{2d}$ to enforce consistency between 2D student and teacher}$
\State $\text{models as per Eq.~(\ref{eq2d}).}$
\State $\text{2. Once trained, freeze the weights of 2D models.}$
\State $\textbf{3D stage:}$
\State $\text{1. Augment 3D student and teacher models with feature}$
\State $\text{maps $F^\text{s2d}_\text{seg}, F^\text{s2d}_\text{sdf}$ and segmentation probability map $\hat{Y}^\text{s2d}$}$
\State $\text{produced by the frozen 2D student model.}$
\State $\text{2. Train the 3D student model with the combined 3D loss}$
\State $\text{$L_\text{3d}$ to enforce consistency between 3D student and teacher}$
\State $\text{models as per Eq.~(\ref{eq3d}).}$
\State $\text{3. Once trained, unfreeze the weights of 2D models.}$
\State $\textbf{Hybrid stage:}$
\State $\text{1. Augment 3D student and teacher models with feature}$
\State $\text{maps $F^\text{s2d}_\text{seg}, F^\text{s2d}_\text{sdf}$ and segmentation probability map $\hat{Y}^\text{s2d}$}$
\State $\text{produced by the trainable 2D student model.}$
\State $\text{2. Calculate uncertainty-weighted hybrid prediction $\hat{Y}^\text{h}$}$
\State $\text{and $\hat{Z}^\text{h}$ from outputs of 2D and 3D teacher models using}$
\State $\text{Eqs.~(\ref{eq5}) and (\ref{eq6}).}$
\State $\text{3. Jointly fine-tune 2D and 3D student models with the}$
\State $\text{combined hybrid loss $L_\text{h}$ to enforce consistency between}$
\State $\text{both student models and hybrid predictions as per Eq.~(\ref{eq10}).}$
\end{algorithmic}
\label{algorithm1}
\end{algorithm}

\begin{figure*}
\centerline{\includegraphics[width=\textwidth]{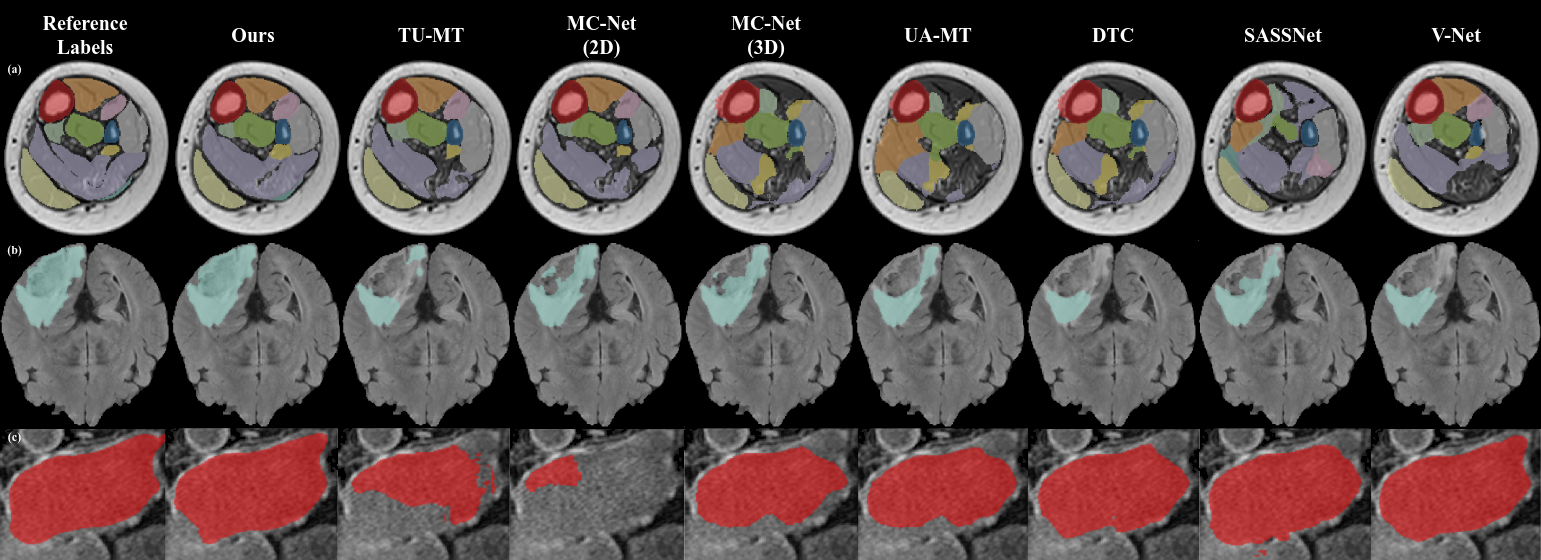}}
\caption{Visual comparison of results produced by the benchmarked networks and the corresponding reference labels. The rows show the results on (a) the MUGgLE dataset, (b) the BraTS2019 dataset, and (c) the LA dataset. From left to right, each row shows the manual reference labels, the segmentation result by our HD-Teacher network, the results by the competing semi-supervised networks TU-MT, MC-Net (2D), MC-Net (3D), UA-MT, DTC, and SASSNet, and finally, the result by the baseline network V-Net trained in full-supervision with all annotated training scans.}
\label{fig_results_all}
\end{figure*}

\section{Experiments \& Results}
\subsection{Dataset Configurations}
HD-Teacher was evaluated on the Left Atrium (LA) dataset \cite{LA}, the Muscle Growth in the Lower Extremity (MUGgLE) dataset \cite{hybridnet2}, and the Multimodal Brain Tumor Segmentation Challenge 2019 (BraTS2019) dataset \cite{BRATS}. All three are MRI datasets and cover different anatomies and have different attributes, enabling a comprehensive evaluation of the proposed method. Scans across all datasets were normalized to zero mean and unit variance to accelerate the training process \cite{u2, flair_ref_2, SASSNet}.

The LA dataset contains 100 3D gadolinium-enhanced MR imaging scans (GE-MRIs) and corresponding manual labels of the left atrial cavity with an isotropic resolution of $0.625 \times 0.625 \times 0.625$ mm. All scans were cropped around the heart region into volumes of $200 \times 200 \times 88$ voxels. We randomly split the dataset into 70 scans for training, 10 for validation, and 20 for testing.

The MUGgLE dataset contains 48 $\text{T}_{1}$-weighted MRI scans of the lower leg of children with cerebral palsy and typically developing children. Each scan was annotated with 13 segmentation labels, representing 11 muscles and two bones of the lower-leg. All scans have an anisotropic resolution of $0.4 \times 0.4 \times 5$ mm. Five scans were annotated by a second rater to determine inter-rater reliability (first row of the middle section of Table \ref{table_all}). All scans were center cropped into volumes of $378 \times 378 \times 96$ voxels. We randomly split the dataset into 35 scans for training, 2 for validation, and 11 for testing.

The BraTS2019 dataset consists of 335 isotropic ($1 \times 1 \times 1$ mm) scans of brain tumors with four modalities available for each ($\text{T}_{1}$, $\text{T}_{1\text{ce}}$, FLAIR, and $\text{T}_{2}$). The FLAIR modality and the whole tumor case were chosen as per \cite{flair_ref, flair_ref_2} for performance evaluation because they provide sufficient information for physicians to investigate malignant tumors. All scans were center cropped into volumes of $200 \times 200 \times 140$ voxels. The dataset was randomly split into 245 scans for training, 25 for validation, and 65 for testing.

\subsection{Experimental Settings \& Evaluation Metrics} \label{lr_decay}
Two experiments were conducted on each dataset using a random selection of 10-11\% and 20\% of the training set, respectively, as the labeled set $D^\text{l}$ to train each network. The rest of the training set was used as the unlabeled set $D^\text{u}$. Both $D^\text{l}$ and $D^\text{u}$ were fixed throughout each experiment so that training data remained the same.

Training was performed with random patches ($112 \times 112 \times 80$ voxels for the LA dataset, $256 \times 256 \times 32$ voxels for the MUGgLE dataset, and $96 \times 96 \times 96$ voxels for the BraTS2019 dataset) with random on-the-fly flip and rotation for augmentation. The stochastic gradient descent (SGD) optimizer was used to train HD-Teacher for 3,000 epochs (for each training stage) with a learning rate decay policy that reduced the learning rate by a factor of 0.1 every 1,000 epochs. The initial learning rate was set to 0.1 in the 2D and 3D stages and to 0.01 in the hybrid stage. The batch size was set to 32 and 2 for the 2D and 3D student models, respectively. Half of each mini-batch was filled with labeled data and the other half with unlabeled data. The EMA decay coefficient $\tau$ was set to 0.99 to control the update to the teacher model. $K$ was set to 8 for the Monte Carlo Dropout. $\lambda^\text{2d}$ and $\lambda^\text{3d}$ in Eq. \eqref{eq10} were set using a time-dependent Gaussian warm-up function $\lambda(i) = 0.1\exp({-5(1 - \frac{i}{i_{\max}})^2})$, where $i$ and $i_{\max}$ denote the current training step and the maximum number of training steps. Hyperparameter $\alpha$ was empirically set to 1 for experiments on the (anisotropic) MUGgLE dataset to utilize more 2D features on anisotropic scans, and to 0.5 for experiments on the (isotropic) datasets LA and BraTS2019. The output of the 3D teacher model was used for inference and performance evaluations. All the benchmarked methods were trained with their original training settings.

Dice Similarity Coefficient (Dice), Jaccard Index (Jaccard), 95\% Hausdorff Distance (HD95), and Average Symmetric Surface Distance (ASD) were used to quantitatively evaluate network performance as per \cite{MTL1, MTL2, DTC, SASSNet}. HD95 and ASD were measured in mm and lower measurements indicate better performance. Dice and Jaccard have a range of $[0, 1]$ and higher measurements represent higher segmentation accuracy. Wilcoxon singed-rank test was used to determine if results between two networks were significantly different. The significance threshold $\alpha$ was set to 0.05.

All experiments were run on a computational node provided by National Computational Infrastructure (NCI) Australia. The node runs the CentOS8 operating system and was configured to have two Nvidia Tesla V100 GPUs, 24 Intel Xeon `Cascade Lake' processors, and 32 GB of RAM.

\subsection{Quantitative \& Qualitative Evaluations}
We benchmarked HD-Teacher against similar consistency regularization-based state-of-the-art semi-supervised methods, including Triple-Uncertainty-Guided Mean-Teacher (TU-MT) \cite{MTL1, MTL2}, 2D and 3D versions of the Mutual Consistency Net (MC-Net) \cite{mcnet1, mcnet2}, Uncertainty-Aware Mean-Teacher (UA-MT) \cite{u2}, Dual Task Consistency Net (DTC) \cite{DTC}, and Shape Aware Semi-Supervised Network (SASSNet) \cite{SASSNet}. The V-Net \cite{VNet} was also trained with all available training data to serve as a reference to the baseline performance achievable with full supervision.

\subsubsection{Results on the LA Dataset}HD-Teacher performed well in both experiments conducted on the LA dataset (see the left section of Table \ref{table_all}). In the first experiment, where seven labeled and 63 unlabeled scans were randomly sampled and fixed for training all networks, HD-Teacher outperformed other semi-supervised methods by at least 3\% Dice and demonstrated a performance very close to the fully-supervised V-Net trained using 70 labeled scans. The second experiment, where 14 labeled and 56 unlabeled scans were randomly sampled and fixed for training, saw HD-Teacher outperforming other semi-supervised methods by a bigger margin ($\ge$ 5\% Dice) and matching the performance of V-Net. Those improvements demonstrated the efficacy of our HD-Teacher. Specifically, the hybrid learning approach allowed HD-Teacher to assign more weights to inter-slice features for the segmentation of isotropic scans, resulting in label maps that accurately capture the 3D structure of the anatomy of interest. By contrast, 2D networks such as TU-MT and MC-Net (2D) may produce under-segmented label maps due to their inability to correlate features from adjacent slices (see the last row of Fig.~\ref{fig_results_all}, columns 2-4). 3D networks such as MC-Net (3D), UA-MT, DTC, and SASSNet produced label maps with reasonable 3D structure yet lacking details, such as some line segments were not correctly curved inward. On the other hand, HD-Teacher selectively merged detailed intra-slice semantic information from the 2D space with inter-slice features and produced label maps that were more detailed than those from 3D networks while retaining structural correctness for the anatomy of interest.

\subsubsection{Results on the MUGgLE Dataset}Similar advantages of HD-Teacher could be observed in experiments conducted on the MUGgLE dataset, which contains anisotropic scans. HD-Teacher and 2D networks performed well in both experiments, while the opposite was true for 3D networks. Specifically, in the first experiment where a fixed set of randomly sampled scans, four labeled and 31 unlabeled, were utilized for training all networks, 2D networks, namely, TU-MT and MC-Net (2D), matched or exceeded the performance of the fully-supervised V-Net trained using 35 labeled scans. 3D networks such as DTC, SASSNet, MC-Net (3D), and UA-MT performed poorly. In comparison, HD-Teacher outperformed all networks by at least 3\% Dice (see middle section of Table 1). The second experiment, where training was conducted using a fixed set of seven labeled and 28 unlabeled scans, showed a similar picture: HD-Teacher demonstrated the best performance, followed by 2D and 3D networks. We attribute the performance of HD-Teacher to its hybrid learning and regularization modules. MC-Net (2D) and TU-MT imposed consistency constraints by minimizing model uncertainty in the 2D space. While functional on anisotropic scans, their inability to capture inter-slice correlation resulted in label maps showing under/over-segmented anatomies with irregular contours (see the first row of Fig.~\ref{fig_results_all}, columns 3 and 4). Similarly, 3D networks enforced consistency or shape constraints in the 3D space. However, the large spacing between slices of anisotropic scans from the MUGgLE dataset prevented effective 3D feature learning, resulting in segmentation results of inadequate quality. HD-Teacher addressed the limitations of 2D and 3D networks via the hybrid learning approach to merge 2D and 3D features for the segmentation task. The proposed hybrid regularization module further imposed model consistency in the hybrid space and dynamically adjusted the amount of 2D/3D information used for training. In the case of anisotropic scans where informative 3D features were difficult to acquire, HD-Teacher diverted more weight to 2D features while still mining spatial context from 3D features, thereby producing segmentation label maps with accuracy that is on par with human expert raters (see the first and last row of the middle section of Table \ref{table_all}).

\subsubsection{Results on the BraTS2019 Dataset}Finally, HD-Teacher also demonstrated the best performance on the BraTS2019 dataset: it outperformed other competing semi-supervised methods in both experiments by a notable margin (see the right section of Table \ref{table_all}). Specifically, our method outperformed all competing methods by a significant margin, including the fully-supervised V-Net, in the first experiment (labeled/unlabeled: 24/221). The increased number of annotated scans in the second experiment (labeled/unlabeled: 49/196) allowed HD-Teacher to improve its performance further. The second and third best results, produced by MC-Net (2D) and TU-MT, respectively, were very close to ours. Nonetheless, the statistical test revealed that our results were significantly different from those of MC-Net (2D) and TU-MT when measured by at least two evaluation metrics. We note that HD-Teacher produced the most visually pleasing label maps with almost no under/over-segmented areas. In comparison, other methods, 2D or 3D, semi- or fully-supervised alike, all produced label maps with either disconnected blobs of pixels or under-segmented anatomies (see the middle row of Fig.~\ref{fig_results_all}). The BraTS2019 dataset has the highest number of slices across all three datasets, yet only a fraction of them contain the anatomy of interest. Those blank slices may distract networks from learning discriminative features. HD-Teacher, with its dual-dimensionality approach, could capture discriminative features from multiple viewpoints to produce accurate label maps.

\begin{table*}[!t]
\centering
\resizebox{1\textwidth}{!}{
\begin{tabular}{lcllllccllllccllll}
\hline
& \multicolumn{5}{c}{LA}&&\multicolumn{5}{c}{MUGgLE}&&\multicolumn{5}{c}{BraTS2019}\\
\cline{2-6}\cline{8-12}\cline{14-18}
& $\boldsymbol{N/M}$ & \bf Dice & \bf Jaccard & \bf HD95 & \bf ASD && $\boldsymbol{N/M}$ & \bf Dice & \bf Jaccard & \bf HD95 & \bf ASD && $\boldsymbol{N/M}$ & \bf Dice & \bf Jaccard & \bf HD95 & \bf ASD\\
\hline
Inter-Rater&\multicolumn{5}{c}{N/A}&&5/0 & 0.86 & 0.77 & 4.9 & 0.9&&\multicolumn{5}{c}{N/A}\\
\hline
VNet        &70/0 &  0.91 & 0.83 & 6.4& 1.5 &&35/0 & 0.79*$\dagger$ & 0.67*$\dagger$ & 7.0*$\dagger$ & 1.3$\dagger$&&245/0& 0.80*$\dagger$ & 0.70*$\dagger$ & 11.4*$\dagger$ & 4.8*$\dagger$\\
\hline
UA-MT       &7/63   & 0.79* & 0.69* & 16.8* & 5.4* &&4/31 & 0.40* & 0.30* & 37.2* & 14.5*&&24/221 & 0.77* & 0.67* & 13.7* & 4.4* \\
SASSNet     &7/63   & 0.86* & 0.76* & 13.1* & 3.3* &&4/31 & 0.40 & 0.29 * & 34.2* & 13.1* &&24/221  & 0.82* & 0.72* & 10.2* & 2.4* \\
DTC         &7/63   & 0.85* & 0.76* & 10.9* & 3.0* &&4/31 & 0.42* & 0.30* & 29.4* & 9.2* &&24/221   & 0.77* & 0.66* & 14.6* & 4.1* \\
MC-Net (2D) &7/63  & 0.64* & 0.52* & 24.9* & 3.6* &&4/31 & 0.80* & 0.70* & 5.4* & 1.3* &&24/221 & 0.79* & 0.69* & 10.6* & 1.6* \\
MC-Net (3D) &7/63   & 0.81* & 0.71* & 14.7* & 2.5* &&4/31 & 0.41* & 0.31* & 39.0* & 16.3* &&24/221& 0.82* & 0.72 & 10.7* & 2.7* \\
TU-MT  &7/63   & 0.83* & 0.71* & 10.1* & 3.2* &&4/31 & 0.79* & 0.68* & 10.8* & 4.1* &&24/221& 0.81* & 0.71* & 9.9* & 1.7* \\
Proposed &7/63 & \textbf{0.89} & \textbf{0.80} & \textbf{7.6} & \textbf{1.9} &&4/31 & \textbf{0.83} & \textbf{0.72} & \textbf{5.2} & \textbf{1.0}&&24/221& \textbf{0.85} & \textbf{0.75} & \textbf{7.6} & \textbf{1.4}\\
\hline
UA-MT       &14/56   & 0.79$\dagger$ & 0.69$\dagger$ & 16.8$\dagger$ & 5.4$\dagger$ &&7/28 & 0.40$\dagger$& 0.30$\dagger$ & 35.5$\dagger$ & 14.0$\dagger$  &&49/196& 0.77$\dagger$ & 0.67$\dagger$ & 13.4$\dagger$ & 3.7$\dagger$ \\
SASSNet     &14/56    & 0.86$\dagger$ & 0.76$\dagger$ & 10.6$\dagger$ & 2.1$\dagger$ &&7/28 & 0.41$\dagger$ & 0.30$\dagger$ & 25.4$\dagger$ & 8.8$\dagger$ &&49/196& 0.83$\dagger$ & 0.73$\dagger$ & 10.0$\dagger$ & 2.6$\dagger$ \\
DTC         &14/56  & 0.85$\dagger$ & 0.76$\dagger$ & 10.3$\dagger$ & 1.8$\dagger$  &&7/28 & 0.42$\dagger$ & 0.31$\dagger$ & 24.7$\dagger$ & 8.4$\dagger$ &&49/196& 0.79$\dagger$ & 0.69$\dagger$ & 13.8$\dagger$ & 4.0$\dagger$ \\
MC-Net (2D) &14/56 & 0.84$\dagger$ & 0.73$\dagger$ & 10.6$\dagger$ & 3.0$\dagger$ &&7/28 & 0.81$\dagger$ & 0.71$\dagger$ & 5.1$\dagger$ & 1.1$\dagger$ &&49/196& 0.85$\dagger$ & 0.76$\dagger$ & 8.2 & 2.0$\dagger$ \\
MC-Net (3D) &14/56 & 0.90 & 0.81 & 7.4 & 1.8 &&7/28 & 0.41$\dagger$ & 0.31$\dagger$ & 39.0$\dagger$ & 16.3$\dagger$ &&49/196& 0.82$\dagger$ & 0.72$\dagger$ & 9.3$\dagger$ & 2.8$\dagger$ \\
TU-MT  &14/56  & 0.86$\dagger$ & 0.76$\dagger$ & 9.5$\dagger$ & 2.4$\dagger$ &&7/28 & 0.81$\dagger$ & 0.70$\dagger$ & 10.3$\dagger$ & 4.0$\dagger$ &&49/196& 0.84$\dagger$ & 0.74$\dagger$ & 8.1 & 1.6 \\
Proposed &14/56 & \textbf{0.91} & \textbf{0.83} & \textbf{6.2} & \textbf{1.6} &&7/28 & \textbf{0.85} & \textbf{0.76} & \textbf{4.4} & \textbf{0.8}&&49/196& \textbf{0.86} & \textbf{0.77} & \textbf{7.5} & \textbf{1.4}\\
\hline
\end{tabular}
}
\caption{Performance of the evaluated methods on all three datasets. The asterisk and dagger symbols denote the statistical significance (p < 0.05) between a network and HD-Teacher trained with the first and second experimental settings. $N/M$ denotes the ratio of labeled to unlabeled scans. Best results are marked in bold.}
\label{table_all}
\end{table*}

\subsection{Ablation Study}
To further investigate the efficacy of HD-Teacher, we conducted an ablation study on the LA and MUGgLE datasets with the first experimental setting (i.e., around $\text{10}\%$ and $\text{90}\%$ of the training set were used as labeled and unlabeled data, respectively) to showcase the improvement each component introduced. The LA dataset provided insight into isotropic datasets with one target class and relatively sufficient reference labels, while the MUGgLE dataset was representative of anisotropic datasets with limited annotated data and multiple classes for segmentation.

\begin{table}[!b]
\centering
\begin{tabular}{lcccc}
\hline
\bf Components & \bf Dice & \bf Jaccard & \bf HD95 & \bf ASD \\
\hline
\multicolumn{5}{c}{LA} \\
\hline
2D Net+2D Reg & 0.77 & 0.67 & 12.6 & 3.1\\
2D Net+SDF+2D Reg & 0.81 & 0.70 & 10.9 & 3.0\\
2D+3D Nets+SDF+Separate Reg&0.87 & 0.78 & 8.0 & 2.0\\
2D+3D Nets+SDF+Hybrid Reg&\textbf{0.89} & \textbf{0.80} & \textbf{7.6} & \textbf{1.9}\\
\hline
\multicolumn{5}{c}{MUGgLE} \\
\hline
2D Net+2D Reg & 0.79 & 0.68 & 6.2 &1.6\\
2D Net+SDF+2D Reg & 0.80 & 0.69 & 6.0 &1.5\\
2D+3D Nets+SDF+Separate Reg&0.82 &0.70 & 5.3 & 1.2\\
2D+3D Nets+SDF+Hybrid Reg&\textbf{0.83} & \textbf{0.72} & \textbf{5.2} & \textbf{1.0}\\
\hline
\end{tabular}
\caption{Ablation study of HD-Teacher on the LA and MUGgLE datasets. Best results are marked in bold.}
\label{table_ablation}
\end{table}

We started with a baseline 2D mean-teacher network with uncertainty-guided regularization (2D Net+2D Reg). We then gradually built our HD-Teacher by adding each component to the baseline model, including multi-task learning (2D Net+SDF+2D Reg), hybrid learning but with separate 2D and 3D uncertainty-guided regularization (2D+3D Nets+SDF+Separate Reg), and finally, hybrid prediction and regularization (2D+3D Nets+SDF+Hybrid Reg).

We observed that all proposed modules consistently improved the segmentation performance on both datasets to varying degrees (Table \ref{table_ablation}). Specifically, on the isotropic LA dataset, adding a supplementary SDF prediction enforced the shared encoder to learn features informative for both segmentation and SDF prediction, thereby allowing the segmentation decoder to be `shape-aware' and produce label maps with refined contours. As a result, the performance was improved by up to 4\% Dice. With the introduction of hybrid learning came a more significant performance gain. Hybrid learning coordinated 2D and 3D feature learning and combined complementary 2D and 3D features into hybrid features. The constructed hybrid features were rich in spatial contexts and intra-slice semantic representations, allowing produced label maps to reflect anatomies accurately with a complete 3D structure. However, under the naive hybrid learning approach, results from 2D and 3D mean-teacher networks were separately regularized with uncertainty estimated from their respective space. Hybrid regularization improved the hybrid approach by first assessing the contributions of 2D and 3D features by merging 2D and 3D teacher predictions into a unified hybrid prediction based on their uncertainty scores. Then, the hybrid uncertainty was estimated from the hybrid prediction, allowing HD-Teacher to impose uncertainty-weighted model consistency in the hybrid space. As a result, a further gain of 2\% Dice was obtained. All modules also improved the performance on the anisotropic MUGgLE dataset, albeit to a lesser degree due to the spatial anisotropy of the scans and the lower number of available labeled training data (4 with MUGgLE and 7 with LA). Nonetheless, each proposed module could improve the model performance consistently, demonstrating their effectiveness further.




\section{Discussion}
Despite the recent progress of semi-supervised networks based on the mean-teacher architecture and uncertainty estimation \cite{u2, MTL1, MTL2} in medical image segmentation, one issue remained unattended to: those networks rely on features learned from images of a single dimensionality to produce segmentation labels and estimate uncertainties from outputs produced by the teacher model to help regularize the student model. This approach is weak, however, as features obtained from a single dimensionality may not cover the full spectrum of information needed for accurate predictions of segmentation labels and uncertainty estimation. For instance, 3D mean-teacher based networks such as the UA-MT could not approximate adequate uncertainty maps to guide their training, leading to an unsatisfactory performance on the MUGgLE dataset. This poses a challenge for mean-teacher networks, as the teacher model may mislead the student model on data where limited information could be extracted by a single 2D/3D model, resulting in reduced performance and generalization.

In response, we developed HD-Teacher, where hybrid, semi-supervised, and multi-task learning come together to tackle the segmentation problem. HD-Teacher utilizes hybrid learning with a 2D and a 3D mean-teacher network to fuse 3D context with detailed 2D features to produce accurate segmentations and SDF predictions. By learning from both spaces, HD-Teacher can effectively deal with data having different voxel sizes by dynamically adjusting the amount of information it extracts from both planes to achieve the best accuracy possible. In addition, we proposed to combine uncertainty-weighted outputs from 2D and 3D teacher models into a single hybrid prediction and estimate the hybrid uncertainty from it. The hybrid uncertainty re-calibrates the hybrid prediction to filter out any unreliable information it may contain, leaving only trustworthy knowledge to regularize 2D and 3D student models. Hybrid learning and regularization allows both student models to learn from reliable information dynamically extracted from 2D and 3D dimensions, effectively addressing the issue of the single-dimensionality approach, thereby achieving better performance and generalization than competing methods.

Extensive experiments were conducted on three datasets covering different MRI modalities, voxel sizes, segmentation tasks, and anatomies. HD-Teacher demonstrated significant performance improvement over benchmarked semi-supervised networks on all three datasets, matching the performance of fully-supervised networks and human expert raters. With an efficient feature learning capability, a low requirement for data annotation, reliable generalizability, and consistent performance, our network could potentially be integrated into various clinical applications and large-scale research to automatically produce segmentation labels, reducing the cost of manual labeling. The proposed multi-task learning, hybrid learning, and regularization modules were also systemically evaluated through an ablation study performed on two representative datasets, illustrating their effectiveness further.

While demonstrating superior performance, HD-Teacher had a few issues worth mentioning. Our framework used a simple U-Net as the backbone and element-wise summation to fuse 2D and 3D features to illustrate the efficacy of the proposed approaches. Substituting those simple designs with their more advanced counterparts would likely improve the segmentation accuracy further. The outputs from 2D and 3D teacher models were also weighted using a simple softmax function. While functional in most scenarios, some edge cases may exist where the softmax function would not suffice. Future work will focus on developing more sophisticated feature learning and weighting mechanisms to capture the information and overall uncertainty from the hybrid space more effectively.

\section{Conclusion}
In this paper, we integrated hybrid, semi-supervised, and multi-task learning into one framework to tackle the challenge of medical image segmentation. The presented dual mean-teacher network utilized 2D and 3D mean-teacher networks to produce hybrid features that led to accurate segmentation labels across a range of MRI scans. A hybrid scheme was also proposed to dynamically combine results from 2D and 3D teacher models for cross-dimensionality regularization, allowing further refinement of the produced segmentation labels. Quantitative and qualitative evaluations on three datasets demonstrated the effectiveness of our approach in binary and multi-class segmentation when limited annotated labels were available for training, revealing its potential for clinical applications due to its human-level performance.


\begin{thebibliography}{00}

\bibitem{segmentation_ref_1}E. Smistad, T. L. Falch, M. Bozorgi, A. C. Elster, and F. Lindseth, ``Medical image segmentation on GPUs – A comprehensive review," {\em Med. Image Anal.}, vol. 20, no. 1, pp. 1-18, 2015.

\bibitem{flair_ref}R. Zeineldin, M. E. Karar, J. Coburger, C. R. Wirtz, and O. Burgert, ``DeepSeg: deep neural network framework for automatic brain tumor segmentation using magnetic resonance FLAIR images,'' {\em Int. J. Comput. Assist. Radiol. Surg.}, vol. 15, no. 6, pp. 909-920, Jun. 2020.

\bibitem{flair_ref_2}X. Luo \emph{et al.,}  ``Semi-supervised medical image segmentation via uncertainty rectified pyramid consistency,'' {\em Med. Image Anal.}, vol. 80, 2022. Art. no. 102517.

\bibitem{nnUNet}F. Isensee, P. F. Jaeger, S. A. A. Kohl, J. Petersen, and K. H. Maier-Hein, ``nnU-Net: a self-configuring method for deep learning-based biomedical image segmentation,'' {\em Nat. Methods}, vol. 18, no. 2, pp. 203-211, Feb. 2021.

\bibitem{UNet}O. Ronneberger, P. Fischer, and T. Brox, ``U-Net: Convolutional networks for biomedical image segmentation,'' in {\em Proc. MICCAI}, vol. 9351, 2015, pp. 234-241.

\bibitem{VNet}F. Milletari, N. Navab, and S.-A. Ahmadi, ``V-Net: Fully convolutional neural networks for volumetric medical image segmentation,'' in {\em Proc. 4th Int. Conf. 3D Vision}, 2016, pp. 565–571.

\bibitem{interaction_1}G. Wang \emph{et al.,} ``Interactive medical image segmentation using deep learning with image-specific fine tuning,'' {\em IEEE Trans. Med. Imag.}, vol. 37, no. 7, pp. 1562-1573, Jul. 2018.

\bibitem{interaction_2}G. Wang \emph{et al.,} ``DeepIGeoS: A deep interactive geodesic framework for
medical image segmentation,'' {\em IEEE Trans. Pattern Anal. Mach. Intell.}, vol. 41, no. 7, pp. 1559–1572, Jul. 2018.

\bibitem{interaction_3}X. Luo \emph{et al.,} ``MIDeepSeg: Minimally interactive segmentation of unseen objects from medical images using deep learning,'' {\em Med. Image Anal.}, vol. 72, Aug. 2021, Art. no. 102102.

\bibitem{weak_supervision_1}F. Gao \emph{et al.,} ``Segmentation only uses sparse annotations: Unified weakly and semi-supervised learning in medical images,'' {\em Med. Image Anal.}, vol. 80, 2022, Art. no. 102515.

\bibitem{weak_supervision_2}G. Valvano, A. Leo, and S. A. Tsaftaris, ``Learning to segment from scribbles using multi-scale adversarial attention gates,'' {\em IEEE Trans. Med. Imag.}, vol. 40, no. 8, pp. 1990-2001, Aug. 2021.

\bibitem{u1}Y. Wang \emph{et al.,} ``Double-uncertainty weighted method for semi-supervised learning,'' in {\em Proc. MICCAI}, vol. 12261, 2020, pp. 542-551.

\bibitem{u2}L. Yu, S. Wang, X. Li, C. Fu, and P. A. Heng, ``Uncertainty-aware self-ensembling model for semi-supervised 3D left atrium segmentation,'' in {\em Proc. MICCAI}, vol. 11765, 2019, pp. 605-613.

\bibitem{u3}L. Hu \emph{et al.,} ``Semi-supervised NPC segmentation with uncertainty and attention guided consistency,'' {\em Knowl Based Syst.}, vol. 239, 2022, Art. no. 108021.

\bibitem{mean_teacher}A. Tarvainen, and H. Valpola, ``Mean teachers are better role models: Weight-averaged consistency targets improve semi-supervised deep learning results,'' in {\em Proc. NeurIPS}, 2017, pp. 1195–1204.

\bibitem{SASSNet}S. Li, C. Zhang, and X. He, ``Shape-aware semi-supervised 3D semantic segmentation for medical images,'' in {\em Proc. MICCAI}, vol. 12261, 2020, pp. 552-561.

\bibitem{DTC}X. Luo, J. Chen, T. Song, and G. Wang, ``Semi-supervised medical image segmentation through dual-task consistency,'' in {\em Proc. AAAI}, 2021, pp. 8801-8809.

\bibitem{mcnet1}Y. Wu \emph{et al.,} ``Mutual consistency learning for semi-supervised medical image segmentation,'' {\em Med. Image Anal.}, vol. 81, 2022, Art. no. 102530.

\bibitem{mcnet2}Y. Wu, M. Xu, Z. Ge, J. Cai, and L. Zhang, ``Semi-supervised left atrium segmentation with mutual consistency training,'' in {\em Proc. MICCAI}, vol. 12902, 2021, pp. 297-306.

\bibitem{MTL1}K. Wang \emph{et al.,} ``Tripled-Uncertainty guided mean teacher network for semi-supervised medical image segmentation,'' in {\em Proc. MICCAI}, vol. 12902, 2021, pp. 450-460.

\bibitem{MTL2}K. Wang \emph{et al.,} ``Semi-supervised medical image segmentation via a tripled-uncertainty guided mean teacher network with contrastive learning,'' {\em Med. Image Anal.}, vol. 79, 2022, Art. no. 102447.

\bibitem{SDF}Y. Xue \emph{et al.,} ``Shape-aware organ segmentation by predicting signed distance maps,'' in {\em Proc. AAAI}, 2020, pp. 12565-12572.

\bibitem{SDF_2}S. Dangi, C. A. Linte, and Z. Yaniv, ``A distance map regularized CNN for cardiac cine MR image segmentation,'' {\em Med. Phys.}, vol. 46, no. 12, pp. 5637-5651, 2019.

\bibitem{soft_1}H. Guo, B. Pasunuru, and M. Bansal, ``Dynamic multi-level multi-task learning for sentence simplification,'' 2018, {\em arXiv:1806.07304}.

\bibitem{soft_2}Y. Yang, and T. M. Hospedales, ``Trace norm regularised deep multi-task learning,'' 2016, {\em arXiv:1606.04038}.

\bibitem{hard_1}R. Girshick, J. Donahue, T. Darrell, and J. Malik, ``Rich feature hierarchies for accurate object detection and semantic segmentation,'' in {\em Proc. IEEE CVPR}, 2016, pp. 580-587.

\bibitem{multi_learning_1}P. Moeskops \emph{et al.,} ``Deep learning for multi-task medical image segmentation in multiple modalities,'' in {\em Proc. MICCAI}, vol. 9901, 2016, pp. 478-486.

\bibitem{multi_learning_2}Y. Zhou \emph{et al.,} ``Multi-task learning for segmentation and classification of tumors in 3D automated breast ultrasound images,'' {\em Med. Image Anal.}, vol. 70, 2021, Art. no. 101918.

\bibitem{multi_learning_3}J. Cheng, J. Liu, H. Kuang, and J. Wang, ``A fully automated multimodal MRI-based multi-task learning for glioma segmentation and IDH genotyping,'' {\em IEEE Trans. Med. Imag.}, vol. 41, no. 6, pp. 1520-1532, Jun. 2022.

\bibitem{multi_learning_4}J. Duan \emph{et al.,} ``Automatic 3D bi-ventricular segmentation of cardiac images by a shape-refined multi-task deep learning approach,'' {\em IEEE Trans. Med. Imag.}, vol. 38, no. 9, pp. 2151-2164, Sept. 2019.

\bibitem{multi_learning_5}K. He \emph{et al.,} ``HF-UNet: learning hierarchically inter-task relevance in multi-task U-Net for accurate prostate segmentation in CT images,'' {\em IEEE Trans. Med. Imag.}, vol. 40, no. 8, pp. 2118-2128, Aug. 2021.

\bibitem{multi_learning_6}F. Uslu, M. Varela, G. Boniface, T. Mahenthran, H. Chubb and A. A. Bharath, ``LA-Net: a multi-task deep network for the segmentation of the left atrium,'' {\em IEEE Trans. Med. Imag.}, vol. 41, no. 2, pp. 456-464, Feb. 2022.

\bibitem{2.5D}H. Zhang \emph{et al.,} ``Multiple sclerosis lesion segmentation with tiramisu and 2.5D stacked slices,'' in {\em Proc. MICCAI}, vol. 11766, 2019, pp. 338-346.

\bibitem{2.5D_1}X. Han, ``Automatic liver lesion segmentation using a deep convolutional neural network method,'' 2017, {\em arXiv:1704.07239}.

\bibitem{2.5D_2}G. Wang \emph{et al.,} ``Semi-supervised segmentation of radiation-induced pulmonary fibrosis from lung CT scans with multi-scale guided dense attention,'' {\em IEEE Trans. Med. Imag.}, vol. 41, no. 3, pp. 531-542, Mar. 2022.

\bibitem{scse}A. G. Roy, N. Navab, and C. Wachinger, ``Concurrent spatial and channel `squeeze \& excitation' in fully convolutional networks,'' in {\em Proc. MICCAI}, vol. 11070, 2018, pp. 421-429.

\bibitem{dropout}N. Srivastava, G. Hinton, A. Krizhevsky, I. Sutskever, and R. Salakhutdinov, ``Dropout: A simple way to prevent neural networks from overfitting,'' {\em J. Mach. Learn. Res.}, vol. 15, no. 56, pp. 1929-1958, 2014.

\bibitem{p_label}W. Bai \emph{et al.,} ``Semi-supervised learning for network-based cardiac MR image segmentation,'' in {\em Proc. MICCAI}, vol. 10434, 2017, pp. 253-260.

\bibitem{data_con_1}S. Laine, and T. Aila, ``Temporal ensembling for semi-supervised learning,'' 2016, {\em arXiv/1610.02242}.

\bibitem{mc_dropout}A. Kendall, and Y. Gal, ``What uncertainties do we need in Bayesian deep learning for computer vision?'' in {\em Proc. NeurIPS}, 2017, pp. 5580–5590.

\bibitem{h_denseunet}X. Li, H. Chen, X. Qi, Q. Dou, C.-W Fu, and P.-A Heng, ``H-DenseUNet: Hybrid densely connected UNet for liver and tumor segmentation from CT volumes,'' {\em IEEE Trans. Med. Imag.}, vol. 37, no. 12, pp. 2663–2674, Dec. 2018.

\bibitem{hybridnet1}J. Zhu, B. Bolsterlee, B. Chow, Y. Song, and E. Meijering, ``Hybrid attentive Unet for segmentation of lower leg muscles and bones from MRI scans for musculoskeletal research,'' {\em Proc. IEEE ISBI}, 2022, pp. 1-5.

\bibitem{hybridnet2}J. Zhu \emph{et al.,} ``Deep learning methods for automatic segmentation of lower leg muscles and bones from MRI scans of children with and without cerebral palsy,'' {\em NMR Biomed.}, vol. 34, no. 12, pp. e4609, 2021.

\bibitem{hybridnet3}Y. Zhou, W. Huang, P. Dong, Y. Xia and S. Wang, ``D-UNet: a dimension-fusion u shape network for chronic stroke lesion segmentation,'' {\em IEEE/ACM Trans. Comput. Biol. Bioinf.}, vol. 18, no. 3, pp. 940-950, June 2021.

\bibitem{hybridnet4}H. Gu \emph{et al.,} ``A 2D–3D hybrid convolutional neural network for lung lobe auto-segmentation on standard slice thickness computed tomography of patients receiving radiotherapy,'' {\em Biomed. Eng. Online}, vol. 20, no. 1, pp. 94, Sep 2021.

\bibitem{hybridnet_5}W. Gan \emph{et al.,} ``Automatic segmentation of lung tumors on CT images based on a 2D \& 3D hybrid convolutional neural network,'' {\em Brit. J. Radiol.}, vol. 94, no. 1126, pp. 20210038, 2021.

\bibitem{LA}Z. Xiong \emph{et al.,} ``A global benchmark of algorithms for segmenting the left atrium from late gadolinium-enhanced cardiac magnetic resonance imaging'' {\em Med. Image Anal.}, vol. 67, 2020, Art. no. 101832.

\bibitem{BRATS}B. H. Menze \emph{et al.,} ``The multimodal brain tumor image segmentation benchmark (BRATS),'' {\em IEEE Trans. Med. Imag.}, vol. 34, no. 10, pp. 1993-2024, Oct. 2015.

\end{thebibliography}
\end{document}